\documentclass[11pt]{article} 
  
\usepackage{amsmath,amsfonts,amssymb,graphicx}

\newcommand{\leqc}{\mbox{$ \;\stackrel{(c)}{\leq}\; $}}
\newcommand{\leqd}{\mbox{$ \;\stackrel{(d)}{\leq}\; $}}

\newcommand{\eqa}{\mbox{$ \;\stackrel{(a)}{=}\; $}}
\newcommand{\eqb}{\mbox{$ \;\stackrel{(b)}{=}\; $}}

\newcommand{\eqe}{\mbox{$ \;\stackrel{(e)}{=}\; $}}

\newcommand{\PR}{\mbox{\rm Pr}}

\def\be{\begin{eqnarray}}
\def\ee{\end{eqnarray}}
\def\ben{\begin{eqnarray*}}
\def\een{\end{eqnarray*}}

\def\slabel#1{\label{s:#1}}

\def\elabel#1{\label{e:#1}}

\def\sq{$\Box$}

\def\qed{\ifmmode\sq\else{\unskip\nobreak\hfil
\penalty50\hskip1em\null\nobreak\hfil\sq
\parfillskip=0pt\finalhyphendemerits=0\endgraf}\fi\par\medbreak}

\newsavebox{\junk}
\savebox{\junk}[1.6mm]{\hbox{$|\!|\!|$}}

\def\liminf{\mathop{\rm lim\ inf}}

\def\til={{\widetilde =}}

 \def\eq#1/{(\ref{#1})}

\def\eq#1/{(\ref{e:#1})}

\newcommand{\beqn}[1]{\notes{#1}%
\begin{eqnarray} \elabel{#1}}

\newcommand{\eeqn}{\end{eqnarray} }

\newcommand{\beq}[1]{\notes{#1}%
\begin{equation}\elabel{#1}}

\newcommand{\eeq}{\end{equation}} 

\def\bdes{\begin{description}}
\def\edes{\end{description}}

\def\notes#1{}

\begin{document}
  
\title{
Some Information-Theoretic Computations\\
Related to the Distribution of Prime Numbers\\
\vspace*{0.5in}
{\large\sc Submitted to Jorma Rissanen's Festschrift volume}\\
\vspace*{0.5in}
}

\author
{
      	I. Kontoyiannis\thanks{Department of Informatics,
		Athens University of Economics and Business,
		Patission 76, Athens 10434, Greece.
                Email: {\tt yiannis@aueb.gr}.
                Web: {\tt http://pages.cs.aueb.gr/users/yiannisk/}.}
}

\maketitle

\begin{abstract}
We illustrate how elementary information-theoretic ideas 
may be employed to provide proofs for well-known, 
nontrivial results in number theory. Specifically,
we give an elementary and fairly short proof 
of the following asymptotic result,
$$\sum_{p\leq n}\frac{\log p}{p}\sim \log n,
\;\;\;\;\mbox{as}\;n\to\infty,$$
where the sum is over all primes $p$ not exceeding $n$.
We also give finite-$n$ bounds refining the above limit.
This result, originally proved by Chebyshev in
1852, is closely related to the celebrated prime number 
theorem.
\end{abstract}


\thispagestyle{empty}
\clearpage

\section{Introduction}
\slabel{intro}

The significant depth of the connection between information
theory and statistics appears to have been recognized 
very soon after the birth of information theory \cite{shannon:48}
in 1948; a book-length exposition was provided by Kullback 
\cite{kullback-book} already in 1959. In subsequent decades 
much was accomplished, and in the 1980s the development 
of this connection culminated in Rissanen's celebrated work
\cite{rissanen:83}\cite{rissanen:87}\cite{rissanen:book}, 
laying the foundations 
for the notion of stochastic complexity and the Minimum 
Description Length principle, or MDL.

Here we offer a first glimpse of a different
connection, this time between information theory 
and number theory. In particular, we will show that basic 
information-theoretic arguments combined with elementary 
computations can be used to give a new proof
for a classical result concerning the distribution 
of prime numbers. The problem of understanding 
this ``distribution'' (including the issue of exactly 
what is meant by that statement) has, of course, been at
the heart of mathematics since antiquity, and it
has led, among other things,
to the development of the field of analytic 
number theory; 
e.g., Apostol's text \cite{apostol:book}
offers an accessible introduction and 
\cite{bateman-diamond:96} gives a more
historical perspective.

A major subfield is {\em probabilistic} number theory,
where probabilistic tools are used to derive results
in number theory. This approach, pioneered by,
among others, Mark Kac and Paul Erd\"{o}s from the 1930s on,
is described, e.g., in Kac's beautiful book \cite{kac:book:59},
Billingsley's review 
\cite{billingsley:1974}, and Tenenbaum's
more recent text \cite{tenenbaum:book}.
The starting point in much of the relevant literature
is the following setup: 
For a fixed, large integer $n$,
choose a random integer $N$
from $\{1,2,\ldots,n\}$, and write it
in its unique prime factorization,
\be
N=\prod_{p\leq n}p^{X_p},
\label{eq:fta}
\ee
where the product runs over all primes
$p$ not exceeding $n$, and 
$X_p$ is the largest power
$k\geq 0$ such that $p^k$ divides $N$.
Through this representation, the uniform distribution
on $N$ induces a joint distribution on the 
$\{X_p\;;\;p\leq n\}$, and the key observation
is that, for large $n$, the 
random variables
$\{X_p\}$ are distributed
approximately like independent geometrics.
Indeed, since there are exactly
$\lfloor n/p^k\rfloor$ multiples of $p^k$
between 1 and $n$, 
\be
\PR\{X_p\geq k\}=\PR\{N\;\mbox{is a multiple of}\;p^k\}=
\frac{1}{n}\Big\lfloor\frac{n}{p^k}\Big\rfloor
\approx\Big(\frac{1}{p}\Big)^k,\;\;\;\;\mbox{for large}\;n,
\label{eq:pmf}
\ee
so the distribution of $X_p$ is approximately geometric.
Similarly, for the joint
distribution of the $\{X_p\}$ we find, 
$$\PR\{X_{p_i}\geq k_{p_i}\;\mbox{for primes }p_1,p_2,\ldots,
p_m\leq n\}
=\frac{1}{n}\Big\lfloor\frac{n}{p_1^{k_1}p_2^{k_2}\cdots p_m^{k_m}}\Big\rfloor
\approx
\Big(\frac{1}{p_1}\Big)^{k_1}
\Big(\frac{1}{p_2}\Big)^{k_2}
\cdots
\Big(\frac{1}{p_m}\Big)^{k_m},$$
showing that the $\{X_p\}$
are approximately independent.

This elegant approximation is also mathematically powerful,
as it makes it possible to translate standard results about 
collections of independent random variables
into important properties that hold for every 
``typical'' integer $N$.
Billingsley in his 1973 Wald Memorial Lectures
\cite{billingsley:1974} gives an account of the 
state-of-the-art of related results up to that point,
but he also goes on to make a further,
fascinating connection with the {\em entropy} 
of the random variables $\{X_p\}$.

Billingsley's argument essentially begins with the
observation that, since the representation (\ref{eq:fta})
is unique, the value of $N$ and the values of the
exponents $\{X_p\}$ are in a one-to-one correspondence;
therefore, the entropy of $N$ is the same as the
entropy of the collection $\{X_p\}$,\footnote{For 
definiteness, we take $\log$ to denote the
natural logarithm to base $e$ throughout, although
the choice of the base of the logarithm is largely
irrelevant for our considerations.}
$$\log n = H(N) = H(X_p\;;\;p\leq n).$$
And since the random variables $\{X_p\}$ are approximately
independent geometrics, 
we should expect that,
\be
\log n = H(X_p\;;\;p\leq n)\approx \sum_{p\leq n} H(X_p)
\approx 
\sum_{p\leq n}\Big[
\frac{\log p}{p-1}-\log\Big(1-\frac{1}{p}\Big)
\Big],
\label{eq:bill}
\ee
where 
in the last equality we simply substituted
the well-known expression for the entropy
of a geometric random variable (see Section~2 for
details on the definition of the entropy and
its computation).
For large $p$, the above summands behave like
$\frac{\log p}{p}$ to first order, 
leading to the asymptotic estimate,
\ben
\sum_{p\leq n}\frac{\log p}{p}\approx\log n,
\;\;\;\;\mbox{for large}\;n.
\een
Our main goal in this paper is to show that this
approximation can indeed be made rigorous, mostly
through elementary information-theoretic arguments; we will 
establish:

\medskip

\noindent
{\bf Theorem~1.} 
As $n\to\infty$,
\be
C(n):=\sum_{p\leq n}\frac{\log p}{p}\sim \log n,
\label{eq:thm1}
\ee
where the sum is over all primes $p$ not 
exceeding $n$.\footnote{As usual, the
notation ``$a_n\sim b_n$ as $n\to\infty$''
means that $\lim_{n\to\infty}a_n/b_n=1$.}

\medskip

\noindent
As described in more detail in the following section, 
the fact that the joint distribution of the
$\{X_p\}$ is asymptotically close to the 
distribution of independent geometrics is
not sufficient to turn 
Billingsley's heuristic 
into an actual proof -- at least, we were 
not able to make the two ``$\approx$''
steps in (\ref{eq:bill}) rigorous directly.
Instead, we provide a proof in two steps.
We modify Billingsley's heuristic to derive
a {\em lower bound}
on $C(n)$ in Theorem~2, and in Theorem~3 
we use a different argument, 
again going via the entropy of $N$,
to compute a corresponding 
{\em upper bound}. These two combined prove
Theorem~1, and they also give finer, 
finite-$n$ bounds on $C(n)$.


In Section~2 we state our main results and
describe the intuition behind their proofs.
We also briefly review some other elegant 
information-theoretic arguments connected 
with bounds on the number of primes up to $n$.
The appendix contains the remaining proofs.

Before moving on to the results themselves, 
a few words about
the history of Theorem~1 are in order. 
The relationship (\ref{eq:thm1}) was first 
proved by Chebyshev 
\cite{chebyshev:1852a}\cite{chebyshev:1852b}
in 1852, where he also produced finite-$
n$ bounds on $C(n)$, with explicit constants.
Chebyshev's motivation was to prove
the celebrated prime number theorem (PNT),
stating that $\pi(n)$, the number of 
primes not exceeding $n$ , grows like,
\ben
\pi(n)\sim\frac{n}{\log n},
\;\;\;\;\mbox{as}\;n\to\infty.
\een
This was conjectured by
Gauss around 1792, and it was only 
proved in 1896;
Chebyshev was not able to produce a complete proof,
but he used (\ref{eq:thm1}) and his finer bounds on 
$C(n)$ to show that $\pi(n)$ is of order 
$\frac{n}{\log n}$.
Although we will not pursue this 
direction here, it is actually not 
hard to see that 
the asymptotic behavior 
of $C(n)$ is intimately connected with 
that of $\pi(n)$. For example, a simple 
exercise in summation by parts shows
that $\pi(n)$ can be expressed directly
in terms of $C(n)$:
\be
\pi(n)=\frac{n+1}{\log (n+1)}\,C(n)-\sum_{k=2}^n
\Big(
\frac{k+1}{\log(k+1)}-
\frac{k}{\log k}
\Big)
C(k),
\;\;\;\;\mbox{for all}\;n\geq 3.
\label{eq:sumbp}
\ee
For the sake of completeness, this is
proved in the appendix.

The PNT was finally proved in 
1896 by Hadamard and
by de la Vall\'ee-Pousin. Their 
proofs were not elementary -- 
both relied on the use of Hadamard's theory 
of integral functions applied to the 
Riemann zeta function $\zeta(s)$;
see \cite{bateman-diamond:96} for some details.
In fact, for quite some time it was
believed that no elementary proof would ever
be found, and G.H.\ Hardy in a famous lecture to 
the Mathematical Society of Copenhagen in 1921
\cite{bohr:52} went as far as to suggest that
``{\em if anyone produces an elementary
proof of the PNT ...  he will show that ...
it is time for the books to be cast aside 
and for the theory to be rewritten.}''
It is, therefore, not surprising that
Selberg and Erd\"{o}s'
announcement in 1948 that they had
produced such an elementary proof caused 
a great sensation in the mathematical world;
see \cite{diamond:82} for a survey.
In our context, it is interesting to note
that Chebyshev's result is again used
explicitly in one of the steps of
this elementary proof.

Finally we remark that, although the simple
arguments in this work fall short of giving
estimates precise enough for an elementary
information-theoretic proof of the PNT, it
may not be entirely unreasonable to hope that
such a proof may exist.


\section{Primes and Bits: Heuristics and Results}
\slabel{heuristics}

\subsection{Preliminaries}

For a fixed (typically large) $n\geq 2$, our starting point is
the setting described in the introduction. 
Take $N$ to be a uniformly distributed integer in $\{1,2,\ldots,n\}$
and write it in its unique prime factorization as in (\ref{eq:fta}),
$$N
\,=\,
\prod_{p\leq n}p^{X_p}
\,=\,
p_1^{X_1}\cdot p_2^{X_2}\cdot \cdots\cdot p_{\pi(n)}^{X_{\pi(n)}},
$$
where $\pi(n)$ denotes the number of primes 
$p_1,p_2,\ldots,p_{\pi(n)}$ up to $n$,
and $X_p$ is the largest integer power $k\geq 0$
such that $p^k$ divides $N$. As noted in 
(\ref{eq:pmf}) above, the distribution 
of $X_p$ can be described by,
\be
\PR\{X_p\geq k\} = 
\frac{1}{n}\Big\lfloor\frac{n}{p^k}\Big\rfloor.
\;\;\;\;\mbox{for all}\;k\geq 1,
\label{eq:distr}
\ee
This representation also gives simple upper 
and lower bounds on its mean $E(X_p)$, 
\be
\mu_p
&:=&
	E(X_p)\;=\;
	\sum_{k\geq 1}\PR\{X_p\geq k\}
	\leq
	\sum_{k\geq 1}\Big(\frac{1}{p}\Big)^k
	=\frac{1/p}{1-1/p}=\frac{1}{p-1},
	\label{eq:ubmu}\\
\mbox{and}
\;\;\;
\mu_p
&\geq&
	\PR\{X_p\geq 1\}
	\geq
	\frac{1}{p}-\frac{1}{n}.
	\label{eq:lbmu}
\ee


Recall the important observation that the distribution 
of each $X_p$ is close to a geometric. To be precise,
a random variable $Y$ with values in $\{0,1,2,\ldots\}$
is said to have a geometric distribution with mean $\mu>0$,
denoted $Y\sim\mbox{Geom}(\mu)$, if 
$\PR\{Y=k\}=\mu^k/(1+\mu)^{k+1}$,
for all $k\geq 0$.
Then $Y$ of course
has mean $E(Y)=\mu$ and its entropy is,
\be
h(\mu):=H(\mbox{Geom}(\mu))=
-\sum_{k\geq 0}\PR\{Y=k\}\log\PR\{Y=k\}=
(\mu+1)\log(\mu+1)-\mu\log\mu.
\label{eq:hmu}
\ee
See, e.g., \cite{cover:book} for the standard properties
of the entropy.

\subsection{Billingsley's Heuristic and Lower Bounds on $C(n)$}

First we show how Billingsley's heuristic can be modified
to yield a lower bound on $C(n)$. Arguing as in the introduction, 
\be
\log n
\eqa
	H(N)
\eqb
	H(X_p\;;\;p\leq n)
\leqc
	\sum_{p\leq n} H(X_p)
\leqd
	\sum_{p\leq n} H(\mbox{Geom}(\mu_p))
\eqe
	\sum_{p\leq n} h(\mu_p),
\label{eq:bill2}
\ee
where $(a)$ is simply the entropy of the uniform
distribution, $(b)$ comes from the fact that $N$
and the $\{X_p\}$ are in a one-to-one correspondence,
$(c)$ is the well-known subadditivity of the entropy,
$(d)$ is because the geometric has maximal
entropy among all distributions on the nonnegative integers
with a fixed mean, and $(e)$ is the definition of $h(\mu)$
in (\ref{eq:hmu}). Noting that $h(\mu)$ is nondecreasing
in $\mu$ and recalling the upper bound on $\mu_p$ 
in (\ref{eq:ubmu}) gives,
\be
\log n
\leq \sum_{p\leq n}h(\mu_p)
\leq \sum_{p\leq n}h(1/(p-1))
= \sum_{p\leq n}\Big[
\frac{p}{p-1}\log\Big(\frac{p}{p-1}\Big)
-\frac{1}{p-1}
\log\Big(\frac{1}{p-1}\Big)
\Big].
\label{eq:bill3}
\ee
Rearranging the terms in the sum proves:

\medskip

\noindent
{\bf Theorem~2.} 
For all $n\geq 2$,
$$
T(n):=\sum_{p\leq n}
\Big[\frac{\log p}{p-1}-\log\Big(1-\frac{1}{p}\Big)\Big]
\geq 
\log n.$$

\medskip

\noindent
Since the summands above behave like
$\frac{\log p}{p}$ for large $p$, it is not 
difficult to deduce the following lower bounds
on $C(n)=\sum_{p\leq n}\frac{\log p}{p}\;$:

\bigskip

\noindent
{\bf Corollary~1.} {\sc [Lower Bounds on $C(n)$]}
\ben
&(i)&
	\hspace{0.3in}
	\liminf_{n\to\infty}\frac{C(n)}{\log n}\geq 1;\\
&(ii)&
	\hspace{0.3in}
	C(n) \geq \frac{86}{125}\log n-2.35,
	\hspace{0.3in}
	\mbox{for all}\;n\geq 16.
	\hspace{2.0in}
\een

\noindent
Corollary~1 is proved in the appendix.
Part~$(i)$ proves half of Theorem~1,
and~$(ii)$ is a simple evaluation of
the more general bound derived in 
equation (\ref{eq:genLB}) in the
proof: For any $N_0\geq 2$,
we have,
$$C(n)
\geq 
\Big(1-\frac{1}{N_0}\Big)
\Big(1-\frac{1}{1+\log N_0}\Big)
\log n
+C(N_0)-T(N_0),
\;\;\;\;\mbox{for all}\;n\geq N_0.
$$

\subsection{A Simple Upper Bound on $C(n)$}

Unfortunately, it is not clear how to reverse the inequalities
in equations (\ref{eq:bill2}) and (\ref{eq:bill3}) to get a
corresponding upper bound on $C(n)$ -- especially inequality 
$(c)$ in (\ref{eq:bill2}). Instead we use a different argument,
one which is less satisfying from an information-theoretic
point of view, for two reasons. First, although again
we do go via the entropy of $N$, it is not necessary to do so;
see equation (\ref{eq:ub1}) below. And second, 
we need to use an auxiliary result, 
namely, the following rough estimate on the 
sum, $\vartheta(n):=\sum_{p\leq n}\log p$:
\be
\vartheta(n):=\sum_{p\leq n}\log p\leq (2\log 2)n,
\;\;\;\;\mbox{for all}\;n\geq 2.
\label{eq:erdos}
\ee
For completeness, it is proved at the end 
of this section.

To obtain an upper bound on $C(n)$, we note that 
the entropy of $N$,
$H(N)=\log n$, can be expressed in an alternative form:
Let $Q$ denote the probability mass function of $N$, 
so that $Q(k)=1/n$ for all $1\leq k\leq n$.
Since $N\leq n=1/Q(N)$ always, we have,
\be
H(N) = E[-\log Q(N)] \geq E[\log N]
=E\Big[\log\prod_{p\leq n}p^{X_p}\Big]
=\sum_{p\leq n}E(X_p)\log p.
\label{eq:ub1}
\ee
Therefore, recalling (\ref{eq:lbmu}) and 
using the bound (\ref{eq:erdos}),
$$\log n\geq 
\sum_{p\leq n}
\left(
\frac{1}{p}-\frac{1}{n}
\right)\log p
=
 \sum_{p\leq n}\frac{\log p}{p}
-\frac{\vartheta(n)}{n}\geq
 \sum_{p\leq n}\frac{\log p}{p}-2\log 2,
$$
thus proving:


\medskip

\noindent
{\bf Theorem~3.} {\sc [Upper Bound]}
For all $n\geq 2$,
$$\sum_{p\leq n}\frac{\log p}{p}\leq \log n + 2\log 2.$$

\medskip

\noindent
Theorem~3 together with Corollary~1 prove Theorem~1.
Of course the use of the entropy could have 
been avoided entirely: Instead of using that $H(N)=\log n$ in
(\ref{eq:ub1}), we could simply use that $n\geq N$ by
definition, so $\log n\geq E[\log N]$, and proceed as before.

Finally (paraphrasing from \cite[p.~341]{hardy-wright:book})
we give an elegant argument of Erd\"{o}s that employs a cute,
elementary trick to prove the inequality on $\vartheta(n)$ 
in (\ref{eq:erdos}).
First observe that we can restrict attention to odd $n$, since
$\vartheta(2n)=\vartheta(2n-1)$, for all $n\geq 2$
(as there are no even primes other than 2). Let
$n\geq 2$ arbitrary; then every prime $n+1<p\leq 2n+1$ 
divides the binomial coefficient,
$$B:=\binom{2n+1}{n}=\frac{(2n+1)!}{n!(n+1)!},$$
since it divides the numerator but not the denominator,
and hence the product of all these primes also divides $B$.
In particular, their product must be no greater than $B$,
i.e.,
$$
\prod_{n+1<p\leq 2n+1} p
\leq B=
\frac{1}{2}\binom{2n+1}{n}
+\frac{1}{2}\binom{2n+1}{n+1}
\leq\frac{1}{2}(1+1)^{2n+1}
=2^{2n},
$$
or, taking logarithms,
$$
\vartheta(2n+1)-\vartheta(n+1)=
\sum_{n+1<p\leq 2n+1}\log p
=\log\Big[ \prod_{n+1<p\leq 2n+1} p\Big]
\leq (2\log 2)n.
$$
Iterating this bound inductively gives the required result.

\medskip

\subsection{Other Information-Theoretic Bounds on the Primes}

Billingsley in his 1973 Wald Memorial Lectures 
\cite{billingsley:1974} appears to have been the
first to connect the entropy with properties of the
asymptotic distribution of the primes. Although there 
are no results in that work based on information-theoretic 
arguments, he does suggest the heuristic upon which 
part of our proof of Theorem~2 was based, and he
also goes in the opposite direction: He uses probabilistic
techniques and results about the primes to compute the 
entropy of several relevant collections of random variables.

Chaitin in 1979 \cite{chaitin:1979} gave a proof 
of the fact that there are infinitely many primes,
using algorithmic information theory. 
Essentially the same argument proves a slightly 
stronger result, namely that,
$\pi(n)\geq \frac{\log n}{\log\log n +1}$,
for all $n\geq 3$.
Chaitin's proof can easily be translated
into our setting as follows.
Recall the representation (\ref{eq:fta})
of a uniformly distributed integer $N$ in $\{1,2,\ldots,n\}$.
Since $p^{X_p}$ divides $N$, we must have $p^{X_p}\leq n$,
so that each $X_p$ lies
in the range,
$$0\leq X_p\leq \Big\lfloor
\frac{\log n}{\log p}
\Big\rfloor
\leq
\frac{\log n}{\log p}
,$$
and hence,
$H(X_p)\leq 
	\log\big(
	\frac{\log n}{\log p}
	+1
	\big).$
Therefore, arguing as before,
$$\log n
=
	H(N)
=
	H(X_p\;;\;p\leq n)
\leq 
	\sum_{p\leq n}
	H(X_p)
\leq
	\sum_{p\leq n}
	\log\Big(
	\frac{\log n}{\log 2}
	+1
	\Big)
\leq
	\pi(n)(\log\log n+1),
$$
where the last inequality holds for all $n\geq 3.$

It is interesting that the same argument applied 
to a different representation for $N$ yields a 
marginally better bound:
Suppose we write,
$$N=M^2\prod_{p\leq n}p^{Y_p},$$
where $M\geq 1$ is the largest integer such 
that $M^2$ divides $N$, and each of
the $Y_p$ are either zero or one.
Then $H(Y_p)\leq \log 2$ for all $p$,
and the fact that $M^2\leq n$ implies 
that $H(M)\leq\log\lfloor\sqrt{n}\rfloor$.
Therefore,
$$\log n = H(N)=H(M,Y_{p_1},Y_{p_2},\ldots,Y_{p_{\pi(n)}})
\leq H(M)+\sum_{p\leq n}H(Y_p)\leq\frac{1}{2}\log n+\pi(n)\log 2,$$
which implies that $\pi(n)\geq\frac{\log n}{2\log 2}$, for all
$n\geq 2$.

Finally we mention that in Li and Vit\'{a}nyi's text \cite{li-vitanyi:book},
an elegant argument is given for a more accurate lower bound
on $\pi(n)$. Using ideas and results from algorithmic information 
theory, they show that, $\pi(n)=\Omega\big(\frac{n}{(\log n)^2}\big)$.
But the proof (which they attribute to unpublished work by 
P.~Berman (1987) and J.~Tromp (1990)) is somewhat involved,
and uses tools very different to those developed here.

\newpage

\section*{Appendix}
\slabel{proofs}

\noindent
{\sc Proof of the Summation-by-parts Formula (\ref{eq:sumbp}). }
Note that, since $\pi(k)-\pi(k-1)$ is zero unless $k$ is prime,
$C(n)$ can be expressed as a sum over all integers $k\leq n$,
$$C(n)=\sum_{2\leq k\leq n}[\pi(k)-\pi(k-1)]\frac{\log k}{k}.$$
Each of the following steps is obvious, giving,
\ben
\pi(n)
&=&
	\sum_{k=2}^n[\pi(k)-\pi(k-1)]\\
&=&
	\sum_{k=2}^n[\pi(k)-\pi(k-1)]\frac{\log k}{k}\frac{k}{\log k}\\
&=&
	\sum_{k=2}^n
	\Big[
	C(k)-C(k-1)
	\Big]
	\frac{k}{\log k}\\
&=&
	\sum_{k=2}^n
	C(k)
	\frac{k}{\log k}
	-
	\sum_{k=2}^n
	C(k-1)
	\frac{k}{\log k}\\
&=&
	\sum_{k=2}^n
	C(k)
	\frac{k}{\log k}
	-
	\sum_{k=1}^{n-1}
	C(k)
	\frac{k+1}{\log(k+1)}\\
&=&
\frac{n+1}{\log(n+1)}\,C(n)-\sum_{k=2}^n
\Big(
\frac{k+1}{\log(k+1)}-
\frac{k}{\log k}
\Big)
C(k)
\;-\;\frac{2}{\log 2}\,C(1),
\een
as claimed, since $C(1)=0$, by definition.
\qed

\noindent
{\sc Proof of Corollary~1. }
Choose and fix any $N_0\geq 2$ and let $n\geq N_0$ arbitrary. 
Then,
$$
\log n
\leq T(n)
=
T(N_0)
+\sum_{N_0<p\leq n}
\Big[\frac{\log p}{p-1}-\log\Big(1-\frac{1}{p}\Big)\Big]
\leq
T(N_0)
+\sum_{N_0<p\leq n}
\Big[\frac{1}{1-\frac{1}{p}}\frac{\log p}{p}+\frac{N_0}{N_0-1}\frac{1}{p}
\Big],
$$
where the last inequality follows from the inequality
$-\log(1-x)\leq x/(1-\delta)$, for all $0\leq x\leq\delta<1$,
with $\delta=1/N_0$. Therefore,
\be
\log n
&\leq&
T(N_0)
+\sum_{N_0<p\leq n}
\Big[\frac{1}{1-\frac{1}{N_0}}\frac{\log p}{p}+\frac{N_0}{N_0-1}
\frac{\log p}{\log N_0}
\frac{1}{p}
\Big]\nonumber\\
&=&
T(N_0)+\Big(\frac{N_0}{N_0-1}\Big)\Big(
1+\frac{1}{\log N_0}\Big)\Big(C(n)-C(N_0)\Big).
\label{eq:lb}
\ee
Dividing by $\log n$ and letting $n\to\infty$ yields,
$$\liminf_{n\to\infty}\frac{C(n)}{\log n}\geq 
\frac{(N_0-1)\log N_0}{N_0(1+\log N_0)},$$
and since $N_0$ was arbitrary, letting now
$N_0\to\infty$ implies~$(i)$.

For all $n\geq N_0$,
(\ref{eq:lb}) implies,
\be
C(n)
\geq 
\Big(1-\frac{1}{N_0}\Big)
\Big(1-\frac{1}{1+\log N_0}\Big)
\log n
+C(N_0)-T(N_0),
\label{eq:genLB}
\ee
and evaluating this at $N_0=16$ gives $(ii)$.
\qed

\newpage

\section*{Acknowledgments}
Thanks to Peter Harremo\"{e}s for spotting a small
error in an earlier version of this paper.


\bibliographystyle{plain}


\end{document}